\documentclass[10pt]{elsart}

\usepackage{epsfig}
\usepackage{calc}
\usepackage{ifthen}

\def\PgSm{\ifmmode{\rm \Sigma^-}
           \else${\rm \Sigma^-}$\fi}
\def\PgXm{\ifmmode{\rm \Xi^-}
           \else${\rm \Xi^-}$\fi}
\def\Pgpm{\ifmmode{\rm \pi^-}
           \else${\rm \pi^-}$\fi}
\def\Pgpp{\ifmmode{\rm \pi^+}
           \else${\rm \pi^+}$\fi}
\def\Pgpz{\ifmmode{\rm \pi^0}
           \else${\rm \pi^0}$\fi}
\def\PgSp{\ifmmode{\rm \Sigma^+}
           \else${\rm \Sigma^+}$\fi}
\def\PgKm{\ifmmode{\rm K^-}
          \else${\rm K^-}$\fi}
\def\PgXm{\ifmmode{\rm \Xi^-}
           \else${\rm \Xi^-}$\fi}

\def\gevc{\ifmmode\mathrm{ GeV/{\it c}}
          \else$\mathrm{ GeV/{\it c}}$\fi}
\def\gev2c2{\ifmmode\mathrm{\,GeV^2/{{\it c}^2}}
          \else$\mathrm{ GeV^2/{{\it c}^2}}$\fi}
\def\rch{\ifmmode{\langle r^2_{ch} \rangle}
           \else$\langle r^2_{ch} \rangle$\fi}

\def\Q2{\ifmmode Q^2
          \else $Q^2$\fi}

\def\gc2{\ifmmode\mathrm{\,GeV^2/{\it c^2}}
          \else$\mathrm{\,GeV^2/{\it c^2}}$\fi}

\def\fm2{\ifmmode\mathrm{\,fm^2}
          \else$\mathrm{\,fm^2}$\fi}

\def\mum{\ifmmode\,\mu\mathrm{m}
          \else$\,\mu\mathrm{m}$\fi}
\def\murad{\ifmmode\,\mu\mathrm{rad}
          \else$\,\mu\mathrm{rad}$\fi}
\def\mubarn{\ifmmode\,\mu\mathrm{barn}
          \else$\,\mu\mathrm{rad}$\fi}
\def\mrad{\ifmmode\,\mathrm{mrad}
          \else$\,\mathrm{mrad}$\fi}
\def\mm{\ifmmode\,\mathrm{mm}
          \else$\,\mathrm{mm}$\fi}
\def\cm{\ifmmode\,\mathrm{cm}
          \else$\,\mathrm{cm}$\fi}

\begin{document}

\newcounter{hours}%
\newcounter{minutes}%
\newcommand{\printtime}{%
\setcounter{hours}{\time/60}%
\setcounter{minutes}{\time - \value{hours} * 60 }%
\ifthenelse{\value{minutes}<10}{\thehours:0\theminutes}{\thehours:
  \theminutes}}


\begin{frontmatter}

\title{Measurement of the \PgSm\ Charge Radius by \PgSm--Electron
       Elastic Scattering}

\collab{The SELEX Collaboration}

\author[MPI]{I.~Eschrich\thanksref{contact}\thanksref{trc}},
\author[MPI]{H.~Kr\"uger\thanksref{trf}},
\author[MPI]{J.~Simon\thanksref{trl}},
\author[MPI]{K.~Vorwalter\thanksref{trm}},
\author[PNPI]{G.~Alkhazov},
\author[PNPI]{A.G.~Atamantchouk\thanksref{tra}},
\author[ITEP]{M.Y.~Balatz\thanksref{tra}},
\author[PNPI]{N.F.~Bondar},
\author[Fermi]{P.S.~Cooper},
\author[Flint]{L.J.~Dauwe},
% ...
\address[Fermi]{Fermilab, Batavia, IL 60510, U.S.A.}
% ...
\address[ITEP]{Institute of Theoretical and Experimental Physics,
  Moscow, Russia}
\address[MPI]{Max-Planck-Institut f\"ur Kernphysik, 69117 Heidelberg,
  Germany}
% ...
\address[PNPI]{Petersburg Nuclear Physics Institute, St. Petersburg,
  Russia}
% ...
\address[Flint]{University of Michigan-Flint, Flint, MI 48502, U.S.A.}
% ...
\thanks[contact]{Corresponding author. {\em E-mail:} ivo@slac.stanford.edu}
\thanks[trc]{Now at Imperial College, London SW7 2BZ, U.K.}
\thanks[trf]{Present address: Boston Consulting Group, M\"unchen,
  Germany}
\thanks[trl]{Present address: Siemens Medizintechnik, Erlangen, Germany}
\thanks[trm]{Present address: Deutsche Bank AG, Eschborn, Germany}
\thanks[tra]{deceased}
\thanks[trb]{Present address: Infineon Technologies AG, M\"unchen,
  Germany}
\thanks[trp]{Now at University of Karlsruhe, Karlsruhe, Germany}
\thanks[trd]{Now at Instituto de F\'{\i}sica da Universidade Estadual
  de Campinas, UNICAMP, SP, Brazil}
\thanks[trn]{Now at Universit\"at Freiburg, Freiburg, Germany}
\thanks[tre]{Now at Physik-Department, Technische Universit\"at
  M\"unchen, Garching, Germany}
\thanks[tro]{Now at Max-Planck-Institut f\"ur Physik, M\"unchen,
  Germany}
\thanks[trg]{Present address: Lucent Technologies, Naperville, IL}
\thanks[trh]{Now at Instituto de F\'{\i}sica Te\'orica da Universidade
  Estadual Paulista, S\~ao Paulo, Brazil}
\thanks[tri]{Present address: SPSS Inc., Chicago, IL}
\thanks[trj]{Now at University of Alabama at Birmingham, Birmingham,
  AL 35294}
\thanks[trk]{Present address: DOE, Germantown, MD}

\begin{abstract}
  The \PgSm\ mean squared charge radius has been measured
  in the space-like \Q2 range 0.035--0.105~\gev2c2\ by elastic
  scattering of a \PgSm~beam off atomic electrons.
  The measurement was performed with the SELEX~(E781) spectrometer
  using the Fermilab hyperon beam at a mean energy of 610~\gevc.

  We obtain
  $\rch\,_\PgSm=\left(0.61\,\pm 0.12\,(stat.)\:
    \pm 0.09\,(syst.)\right)\,\fm2$.
  The proton and \Pgpm\ charge
  radii were measured as well and are consistent with results of
  other experiments.
  Our result agrees with the recently measured strong interaction
  radius of the \PgSm.
\end{abstract}

\end{frontmatter}

\section{Introduction}
%
The systematic measurement of the static properties of hadrons has
led to a better understanding of their fundamental
structure. However, their finite size -- a consequence of the
confinement of quarks inside a spatial volume -- has not been
thoroughly explored.
Sizes of hadrons may be probed by their strong and electromagnetic
interactions. Most commonly the electromagnetic charge radius
is measured in elastic electron-hadron scattering.
For unstable hadrons the inverse kinematics can be applied 
using a suitable hadronic beam. 
So far, electromagnetic radii have been established only for the
proton \cite{proton}, neutron \cite{kopecky95}, \Pgpm\ 
\cite{amendolia86pion}, and  \PgKm\ \cite{amendolia86kaon}.
The difference between the pion and the kaon radius indicates a
dependence on the strangeness content.
A systematic study of the radii of hyperons with different strangeness
will therefore enhance our understanding of the relative
sizes of hadrons as bound quark systems.
Theoretical predictions for the \PgSm\ charge radius have been
divergent~\cite{povh90}. The most recent
efforts suggest $\rch\,_\PgSm=(0.67 \pm 0.03)\fm2$~\cite{kubis01} on
the one hand and $\rch\,_\PgSm=(0.54 \pm 0.09)\fm2$~\cite{leinweber}
on the other.
The feasibility of probing the \PgSm\ radius by inverse electron
scattering has been demonstrated at CERN~\cite{adamovich99}.
We present here the first high-statistics measurement of the \PgSm\
charge radius. 

\section{The E781/SELEX Experiment}
\begin{figure*}[htb]
\centerline{\psfig{figure=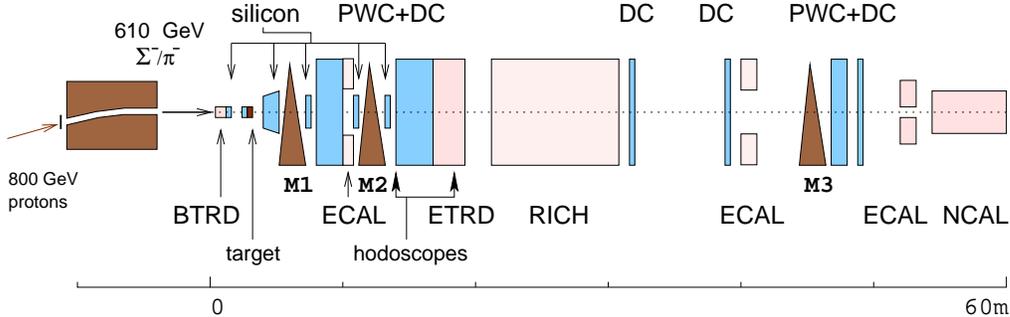,width=\textwidth}}
\caption{Schematic layout of the SELEX apparatus~\cite{layout}.
  The spectrometer
  stages are defined by the three magnets stylized as triangles.
  Transverse dimensions are not to scale.}
\label{fig:selex}
\end{figure*}

%
The primary objective of SELEX is the study of charm hadroproduction and
spectroscopy of charm baryons in the forward hemisphere.
The experiment was built as a 3-stage magnetic spectrometer
as shown in Fig.~\ref{fig:selex}.
Here we only describe those features of the apparatus which are relevant
to hadron-electron scattering.

The 800~\gevc\ proton beam from the Fermilab Tevatron was steered onto
a Be target to produce a beam consisting of equal numbers of
\PgSm\ and \Pgpm\ at 610~\gevc\ mean momentum. The hyperon beam
had a momentum spread of $\Delta p/p=\pm 8\%$.
Other constituents were 1~\% \PgXm\ and less than 0.1~\% other
particles.
The magnet polarity could
be reversed to provide a 540~\gevc\ beam containing 94~\% protons and
3~\% each of \Pgpp\ and \PgSp.
A transition radiation detector (BTRD) separated the baryonic from
the mesonic beam components.
Interactions took place in a target stack of two Cu and three C foils
adding up to 4.2~\% of an interaction length for protons.
The first two magnets (M1, M2) implemented
momentum cuts of 2.5~\gevc\ and 15~\gevc, respectively.
Downstream of the second magnet a second
transition radiation detector (ETRD) identified scattered electrons.
Each stage
of the spectrometer was equipped with a lead glass calorimeter (ECAL).
Tracking information was provided by a combination of silicon
microstrip detectors,
proportional chambers (PWC), and drift chambers (DC). 

Candidates for elastic scattering events were selected by
a scintillator-based trigger requiring two
charged particles in the M2 spectrometer both of which
originated in the target area.
The total multiplicity signal directly downstream of the targets
was combined with the charge and
multiplicity information from the hodoscope after magnet M2. 
The total charge was required to
correspond to one beam particle plus one electron.
An online filter refined this selection by checking for the same
topology using particle tracks reconstructed in the M2 spectrometer.

\section{Kinematics}
%
The differential cross section of the
elastic scattering of a spin 1/2 baryon (mass $M$) off an electron
(mass $m$) can be approximated by the relation ($\hbar=c=1$)\cite{Kal64}
\begin{equation}
  \label{xsecmott}
  \frac{d\sigma}{dQ^2} \, = \, \frac{4\pi\alpha^2}{Q^4} \, 
  \left(1-\frac{Q^2}{Q^2_{max}} \right) \, F^2(Q^2)
\end{equation}
up to corrections of the order $m^2/(s-M^2)$.
Here, $s$ denotes the center of mass energy and $Q^2$ the four momentum
transfer from the hadron to the electron.
\Q2 has a kinematically allowed maximum value which depends on the beam
momentum.
For instance,  $Q^2_{max}$ is 0.2~\gev2c2\ for 610~\gevc\ \PgSm. 
For $m \ll M$ the squared form factor can be written as a combination of
the electric and magnetic form factors $G_E(\Q2)$ and $G_M(\Q2)$:
\begin{equation}
  \label{formfactor}
  F^2(Q^2) \, = \, \frac{G_E^2+({Q^2}/{4M^2})G_M^2}{1+({Q^2}/{4M^2})} 
   + \frac{Q^4}{2(4M^2E^2-(2ME+M^2)Q^2)} G_M^2,
\end{equation}
with $E$ being the beam energy in the laboratory frame.
These form factors are normalized to the charge $G_E(0)=Z$ and the
magnetic moment $G_M(0)=\mu$ of the baryon. The mean squared charge
radius is defined by the relation 
\begin{equation}
  \label{radiusdef}
  \rch \, = \, \frac{-6}{G_E(0)} \, \frac{dG_E}{dQ^2}\Bigg|_{Q^2=0}.
\end{equation}
%
At low momentum transfers $\Q2\leq 0.1 \gev2c2$,
both electric and magnetic form factors
of the \PgSm\
can be parameterized by the dipole approximation:
\begin{equation}
  \label{dipole}
  \frac{G_E(Q^2)}{G_E(0)} = \frac{G_M(Q^2)}{G_M(0)} =   
  D(Q^2) = {\left(1+\frac{\rch Q^2}{12}\right)^{-2}}.
\end{equation}
A fit of equation~(\ref{xsecmott})
to the shape of the measured \Q2\
distribution provides the information on \rch\ by means of
equation~(\ref{dipole}).

\section{Analysis and Results}
%
Out of 77 million triggers with a \PgSm\ a sample of 5010
\PgSm-electron scattering events was extracted as follows:
The events containing one electron were selected and
the particle trajectories combined to check if they formed vertices
inside the target material. The event was accepted if it contained
exactly three tracks
forming one vertex consisting of the beam track,
the electron candidate,
and the scattered hadron candidate.

\begin{figure}[htb]
  \begin{center}
    \leavevmode
    \epsfig{file=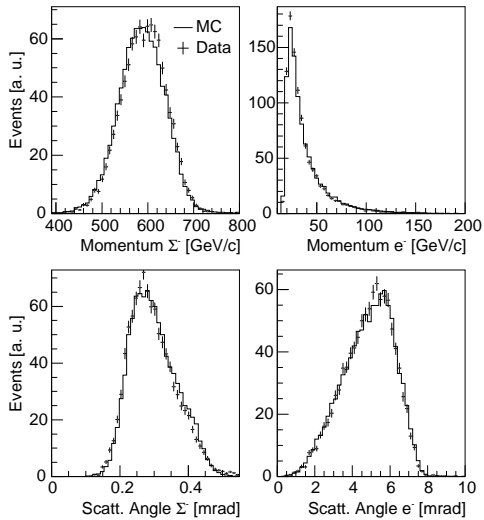,width=7.5cm}
    \caption[]{Comparison of the  elastic scattering observables 
    in data and simulation.
    }
    \label{fig:kinvar}
  \end{center}
\end{figure}

\begin{figure}[htb]
  \begin{center}
    \epsfig{file=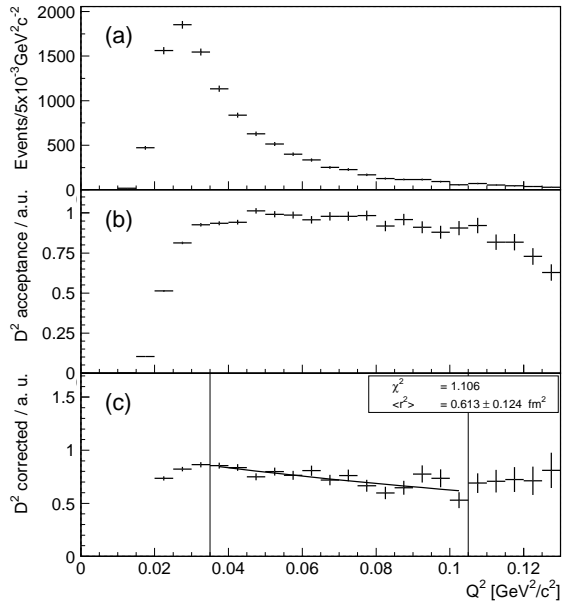,width=7.5cm}
    \caption[]{(a) \Q2 distribution of the final event sample.
      (b) Acceptance function applied to a constant dipole term
      $D^2(Q^2)$.
      (c) Fit to $D^2(Q^2)$ extracted from data
      corrected for acceptance.
      Vertical lines indicate boundaries of fit.
      $D^2$ is given in arbitrary units, statistical error only.}
    \label{fig:dipol_final}
  \end{center}
\end{figure}

At a given beam energy,
momenta and scattering angles of the
outgoing electron and hadron (Fig.~\ref{fig:kinvar})
provide four observables to determine the
four-momentum transfer from hadron to electron
(Fig.~\ref{fig:dipol_final}a).
As the measurement is overdetermined,
the electron momentum could be ignored,
thus avoiding sizable radiative corrections.
This is an important advantage of the reversed method since
in experiments with an electron beam the corrections for
Bremsstrahlung have a direct influence on the result. 
The remaining radiative corrections to the
\Q2 spectrum due to initial Bremsstrahlung loss are small,
and show a variation of less
than 0.2\% in the \Q2\ region used in this analysis.
The momentum transfer was fitted to the remaining three observables.
In the last stage of the event selection,
both scattering angles and the momentum loss of the
hadron were required to satisfy the constraints imposed by elastic
kinematics.

A full GEANT~\cite{geant} simulation of detector, trigger,
reconstruction, and analysis was performed (Fig.~\ref{fig:kinvar}),
which yielded the correction
for acceptance
effects of the experimental setup (Fig.~\ref{fig:dipol_final}b).
At $\Q2<0.03 \gev2c2$ the electron is driven out of the geometric
acceptance as a result of the M2 magnetic field.
For $\Q2>0.11 \gev2c2$ the electron is likely to hit the central
sections of the hodoscopes together with the hadron,
not producing a trigger.
In the remaining \Q2 range the acceptance has only minor variations
reflecting the individual efficiencies of hodoscope segments.

To eliminate kinematic effects from varying beam energy,
the measured differential cross section was weighted with the
Mott scattering cross section, accounting for the magnetic moment
of the hadron.
This yields the squared dipole form factor
$D^2(Q^2)$ which in turn provides \rch
(Eqn.~(\ref{dipole}); Fig.~\ref{fig:dipol_final}c).
This method is more sensitive to \rch\ than an unbinned
maximum likelihood fit of the \Q2\ distribution even though it
adds binning effects to the systematic uncertainty~\cite{sigmaref}.
The fit of $D^2(Q^2)$ was restricted to the \Q2\ region
of minimal systematic error
between 0.035~\gev2c2 and 0.105~\gev2c2.

The systematic errors in \rch\ introduced by cuts on kinematic
variables ($\chi^2$ of the vertex fit, sum of momenta,
scattering angles of electron and hadron) were estimated by
varying one cut at a time by $\pm 1\sigma$ of its resolution.
Systematic dependence of the result on the fit boundaries
was studied over the full \Q2 range. The boundaries were selected
so as to minimize this dependence,
and the remaining contribution to the systematic
error determined by varying them by $\pm 1\sigma$ of the \Q2
resolution.
The contribution of histogram binning effects was evaluated by
systematically changing the bin width.

The total systematic error (table~\ref{tab:systerr}) is dominated 
by the choice of cuts on kinematic variables
on one hand and the choice of fit boundaries on the other.

\begin{table}[htb]
\caption{Contributions to the systematic error of $\rch\,_\PgSm$
  (in \fm2).}
\label{tab:systerr}
\begin{tabular}{lr}
\hline
Cuts on kinematic variables & 0.04\\
Choice of fit boundaries & 0.04\\
Histogram binning effects & 0.03\\
Beam contamination & 0.01\\
\hline
\end{tabular}
\end{table}

The number of events in the final selection
corresponds to 4~\% of
elastic \PgSm\-electron events expected in this interval. 
The efficiency of this measurement was mainly
limited by trigger (23~\%) and reconstruction efficiencies (21~\%).
Background from \PgXm-electron scattering was estimated to be present
in far less than one
percent of the final event sample. A different class of background can
be caused by events
of the form \PgSm~\Pgpz~$\to$~\PgSm~$\gamma e^-e^+$. If the positron
is not detected due to early absorption or tracking inefficiency, the
event topology can appear to be similar to \PgSm-electron scattering.
Events of this kind have been observed
in early SELEX data where the interaction counter had been mistuned.
The invariant mass of reconstructed electrons and photons in the ECAL
shows no evidence of such events in the final event sample.

Our result for the mean squared \PgSm\ electromagnetic radius is
\begin{center}
$\rch\,_\PgSm\;=\;\left(0.61
  \,\pm 0.12\,(stat.)\:\pm 0.09\,(syst.)\right)\,\fm2$. 
\end{center}
In order to facilitate an interpretation of this result free of
any systematics introduced by the experimental setup, 
the mean squared electromagnetic proton radius was measured
as well. Using the same procedure as described for the \PgSm\
analysis~\cite{krueger99} we obtain
\begin{center}
$\rch_p\;=\;\left(0.69\,\pm0.06\,(stat.)\:\pm 0.06\,
  (syst.)\right)\,\fm2$,
\end{center}
which is in good agreement with results from elastic $ep$ scattering
experiments (overall $\rch_p=0.77\pm0.03\,\fm2$~\cite{proton}. 
Ref.~\cite{murphy74} finds $\rch_p=0.67\pm0.02\,\fm2$
from data in a \Q2 range that overlaps well with ours).
In addition we have determined the mean squared \Pgpm\ charge
radius from SELEX data assuming a monopole form factor~\cite{pionref}.
Our result,
\begin{center}
$\rch\,_{\Pgpm}\;=\;\left(0.42\,\pm0.06\,(stat.)\:\pm 0.08\,
  (syst.)\right)\,\fm2$,
\end{center}
is consistent with the results of other experiments
\cite{amendolia86pion}.

\section{Discussion and Conclusion}
\begin{figure}[htbp]
  \begin{center}
    \leavevmode
    \epsfig{file=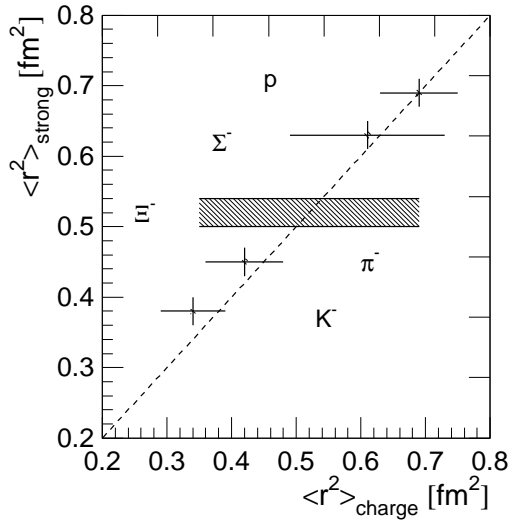,width=7.5cm}
    \caption{Comparison of strong and electromagnetic mean
      squared radii for proton,
      \PgSm, and \Pgpm\ from SELEX data for which only
      statistical errors are displayed.
      Equality of strong interaction and charge radius is indicated
      by the dashed line. The radii of $K^-$ and \PgXm\ from other
      experiments are included for reference (see also
      table~\ref{tab:comprad}).}
    \label{fig:comprad}
  \end{center}
\end{figure}

%
To put our results in perspective,
we compare the electromagnetic charge radii
to radii derived from total hadron-proton cross sections
(strong interaction radii~\cite{povh90}, Fig.~\ref{fig:comprad}).
The latter have been normalized to the mean squared charge radius
of the proton as
measured by SELEX at $\sqrt{s}=34$~GeV and the total $pp$ cross
section~\cite{pdg} interpolated to this energy.
Both the \PgSm\ and \Pgpm\ squared strong interaction radii are
calculated from total cross sections measured by SELEX~\cite{dersch99}
as well. 
The mean squared radii of the proton, \PgSm, and \Pgpm\
were determined using identical experimental setup, reconstruction,
and analysis procedure. 
We therefore expect these particular results to be
free of systematic error in respect to each other,
and only statistical errors are displayed in Fig.~\ref{fig:comprad}.

For reference
we added the squared \PgKm\ charge radius~\cite{amendolia86kaon}
as well as the squared strong interaction radii of \PgKm\ and \PgXm\
to Fig.~\ref{fig:comprad}.
The \PgKm\ strong interaction radius was extrapolated to SELEX
energy using the Particle Data Group's parameterization
of the $K^-p$ total cross section~\cite{pdg}. For the 
\PgXm\ we used the radius determined at
$\sqrt{s}=16$~GeV~\cite{povh87}.
It was not scaled to SELEX energy for lack of
$\PgXm p$ cross section data.

We conclude that our results for the
mean squared electromagnetic charge radii
of proton, \PgSm, and \Pgpm\ agree with the values of their
mean squared strong interaction radii.
They also confirm the charge radii reported
by previous experiments.
The \PgSm\ charge radius is comparable within errors to that of the
proton, yet compatible with a smaller value as suggested by the trend
indicated by strong interaction radii: the results imply a ratio
$\rch_p/\rch_{\PgSm}=1.13\pm0.24$.

\begin{table}[htb]
\caption{Mean squared electromagnetic and strong interaction radii
  shown in Fig.~\ref{fig:comprad}.}
\label{tab:comprad}
\begin{tabular}{lccl}
&Electromagnetic & Strong interaction & \\
& \rch [\fm2] & $\langle r^2_{st} \rangle$ [\fm2] &\\
\hline
$p$    & $0.69\pm0.06$ & $0.69\pm0.02$\\
\PgSm\ & $0.61\pm0.12$ & $0.63\pm0.02$  \\
\Pgpm\ & $0.42\pm0.06$ & $0.45\pm0.02$  \\
$K^-$  & $0.34\pm0.05$ & $0.38\pm0.02$  \\
\PgXm\ &               & $0.52\pm0.02$  \\
\hline
\end{tabular}
\end{table}

Our result for the \PgSm\ is at this point the best direct 
measurement of the charge radius of a charged baryon other
than the proton. It covers an unprecedented range of squared
momentum transfer \Q2 between \PgSm\ and electron.
With this experiment we have proven that it is possible to measure 
baryon charge radii in inverse kinematics with high precision.
Due to the $Q^{-4}$ behavior of the differential cross section
the accuracy of the final result is limited by the low statistics
towards higher \Q2. 

In a dedicated experiment it would be possible to extend
trigger and reconstruction to both lower and higher \Q2,
allowing for increased statistics and better control of the systematics.
By using very finely segmented hodoscopes, for example, one could
trigger on hadron-electron pairs of very small common angle, extending
the acceptance to higher \Q2. Since this relaxes the requirements for
hadron-electron separation one could at the same time reduce
the magnetic field strength downstream
of the targets to have access to lower \Q2 as well.

\ack{The authors are indebted to the staff of Fermi National Accelerator
Laboratory and for invaluable technical support from the staffs of
collaborating institutions, especially F.~Pearsall, D.~Northacker, and
J.~Zimmer.
This project was supported in part by the Bundesministerium f\"ur
Bildung, Wissenschaft, Forschung und Technologie, Consejo
Nacional de Ciencia y Tecnolog\'{\i}a (CONACyT), the Conselho Nacional
de Desenvolvimento Cient\'{\i}fico e Tecnol\'{o}gico, the Fondo de Apoyo
a la Investigaci\'{o}n (UASLP), the Funda\c{c}\~{a}o de Amparo \`{a} 
Pequisa do Estado de S\~{a}o Paulo (FAPESP), the Israel Science
Foundation founded by the Israel Academy of Sciences and Humanities,
Istituto Nazionale di Fisica Nucleare (INFN), the International Science
Foundation (ISF), the National Science Foundation (Phy\#9602178), NATO
(grant CR6.941058-1360/94), the Russian Academy of Science, the Russian
Ministry of Science and Technology, the Turkish Scientific and
Technological Research Board (T\"UB\.ITAK), the U.S. Department of
Energy (DOE grant DE-FG02-91ER40664), and the U.S.-Israel Binational
Science Foundation (BSF).}


\end{document}